\RequirePackage{fix-cm}
\documentclass[twocolumn]{svjour3}          
\smartqed  
\usepackage{graphicx}
\usepackage{amsmath}
\usepackage[ruled]{algorithm2e} 
\usepackage{multirow}
\begin{document}

\title{Hierarchical Cloth Simulation using Deep Neural Networks}
\subtitle{}
\author{}
\institute{}
\author{Young Jin Oh \and Tae Min Lee \and In-Kwon Lee}
\institute{Young Jin Oh \at
			Department of Computer Science, Yonsei University \\
           \email{skrcjstk@gmail.com}
           \and
           Tae Min Lee\at
           	Department of Computer Science, Yonsei University \\
           \email{dnflxoals@yonsei.ac.kr}
           \and
          In-Kwon Lee \at
			Department of Computer Science, Yonsei University \\
           \email{iklee@yonsei.ac.kr}
}
\date{ }

\maketitle

\begin{abstract}
Fast and reliable physically-based simulation techniques are essential for providing flexible visual effects for computer graphics content. In this paper, we propose a fast and reliable hierarchical cloth simulation method, which combines conventional physically-based simulation with deep neural networks (DNN). Simulations of the coarsest level of the hierarchical model are calculated using conventional physically-based simulations, and more detailed levels are generated by inference using DNN models. We demonstrate that our method generates reliable and fast cloth simulation results through experiments under various conditions. 
\end{abstract}

\section{Introduction}
Computer graphics contents, such as movies and games, require fast and reliable physically-based simulation methods for more flexible and realistic visual effects. In particular, in the case of cloth simulation many studies have been continuously undertaken to meet these demands. For example, improved implicit Euler integration~\cite{Bouaziz:2014:PDF:2601097.2601116,Liu:2013:FSM:2508363.2508406,Narain2016} and hierarchical cloth simulation~\cite{BENDER2013945,Lee2010MultiResolutionCS,ZHANG2001383} have generated reasonably fast and accurate simulation results. However, the costs of using these methods remain high, not only in real-time applications, but also in offline graphics systems.

Recently, a type of machine learning technology called a deep neural network (DNN) has been employed in physically-based simulation studies for fast computation. In these studies, DNN has been used to replace complex computational processes with inference processes of trained neural networks~\cite{CNNFluid2016,Yang:2016:DPM:3074919.3074946}.
These studies have shown that DNN can replace complex computational processes and generate reasonable simulation results. However, most studies have focused only on fluid and smoke among the various simulation areas. 
In addition to replacing specific parts of the entire simulation with the inference of DNN, there have been trials in which DNN generates physical motions of rigid body objects, and simple elastic models without using any physically-based simulation~\cite{chang2016compositional,NIPS2017_7040}. These studies have shown that DNN can generate simple physical motions without using any physically-based simulation, but there have also been significant errors such as inaccurate collisions and awkward motions. 

Significant errors resulting from replacing all physically-based simulation result from the inference errors of DNN, which increase cumulatively as the simulation is performed~\cite{chang2016compositional}. Therefore, using DNN for physically-based simulation studies requires a complementary method to generate reliable simulation results.
Moreover, considering the performance of the inference calculations, complex structures such as convolutional neural networks (CNN) and recurrent neural networks (RNN) are not easy to employ. Although these complex DNN structures can be trained more reliably and accurately than simple neural network structures, they are ineffective because of the disadvantage that the inference time is slower than the computation time of conventional physically-based simulation methods. Therefore, we need to use a well-trained DNN that can be accurately inferred for physically-based simulation, but with a simple neural network structure that can be inferred faster than conventional simulation methods.

The hierarchical cloth simulation method proposed in this paper is a hybrid method using both physically-based simulation and DNN. The entire simulation process is divided into two phases. In the first phase, the next state of the coarsest level in the hierarchical model is calculated using a general physically-based simulation method. In the second phase, the next position of finer levels is inferred using DNN models. This inference process using DNN models can be applied continuously to the other higher levels.

Because the proposed method calculates only the coarsest level in the hierarchical cloth model using an accurate physically-based simulation method, it is not accurate compared to the results of conventional physically-based simulation. However, it prevents significant errors stemming from the inference of the DNN model, and guarantees a reliable simulation. Except for the coarsest level calculation, the simple DNN structure used in the inference process has a low-cost inference time, and can be calculated in parallel. Therefore, the total time required for the simulation is faster than the conventional physically-based simulation method. We compare the quality and performance of our simulation results with conventional physically-based simulation methods. Through experiments under various simulation conditions and DNN structures, we show that our method is fast and appropriate for reliable physically-based simulation.

\section{Related Work}
\subsection{Non-linear Elastic Simulation}
A non-linear elastic model is essential to express various elastic materials such as cloth, rubber, muscle, and hair. The integration method of non-linear elastic simulation is important, because it must be able to guarantee stable simulation results. The implicit Euler method~\cite{baraff1998large,Terzopoulos:1987:EDM:37402.37427} can integrate stable simulations even for large time steps, but the traditional implicit Euler method requires solving sparse linear systems at each time step. For fast implicit Euler integration, a solver based on the block coordinate descent scheme for the mass-spring system~\cite{Liu:2013:FSM:2508363.2508406} has been proposed, which converts the implicit Euler integration for the mass-spring system into an energy minimization. Projective dynamics~\cite{Bouaziz:2014:PDF:2601097.2601116} generalized this approach, and has been accelerated through various approaches for the fast solving of the optimization problem~\cite{Liu:2017:QMR:3087678.2990496,Wang:2015:CSA:2816795.2818063,Weiler:2016:PF:2994258.2994282}. Recently, the alternating direction method of multipliers (ADMM) optimization algorithm~\cite{boyd2011distributed} has been applied to the projective dynamics~\cite{Narain2016}, and it has been extended for the handling situations with dynamically changing constraints~\cite{overby2017admmpd}.

On the other hand, a pseudo-physically-based simulation, called position based dynamics (PBD)~\cite{muller2007position}, has been proposed as an alternative approach for the traditional method. Although PBD is limited in areas that require accurate simulation results, it is widely employed in a variety of graphics applications, because it is fast, robust, and easy to implement. Our method uses projective dynamics, applying the ADMM algorithm to calculate the next position of the coarsest level of cloth hierarchy. Because our method only uses the conventional method at the coarsest level, it is not accurate compared with the results of the conventional method. However, our method can generate simulation results quickly compared with the conventional method.

\subsection{Hierarchical Method for Cloth Simulation}
Because high-resolution simulation has been employed for computer graphics contents, hierarchical methods have been proposed for efficient simulation. For fast cloth simulation, the unchanging hierarchical cloth~\cite{ZHANG2001383} and dynamically changing multi-resolution cloth models~\cite{Lee2010MultiResolutionCS} have been proposed. The multi-resolution shape matching method~\cite{BENDER2013945} has been proposed to extend the hierarchical position-based method~\cite{muller2008hierarchical} to cloth simulation. Furthermore, an acceleration method for hierarchical cloth simulation using CPU and GPU concurrently has been proposed~\cite{schmitt2013multilevel}. We use the unchanging hierarchical model for cloth simulation. Unlike previous hierarchical methods, which simulate all levels in the same manner, our method uses both conventional physically-based simulation and the inference of DNN models.

\begin{figure*}
\includegraphics[width=6.8in]{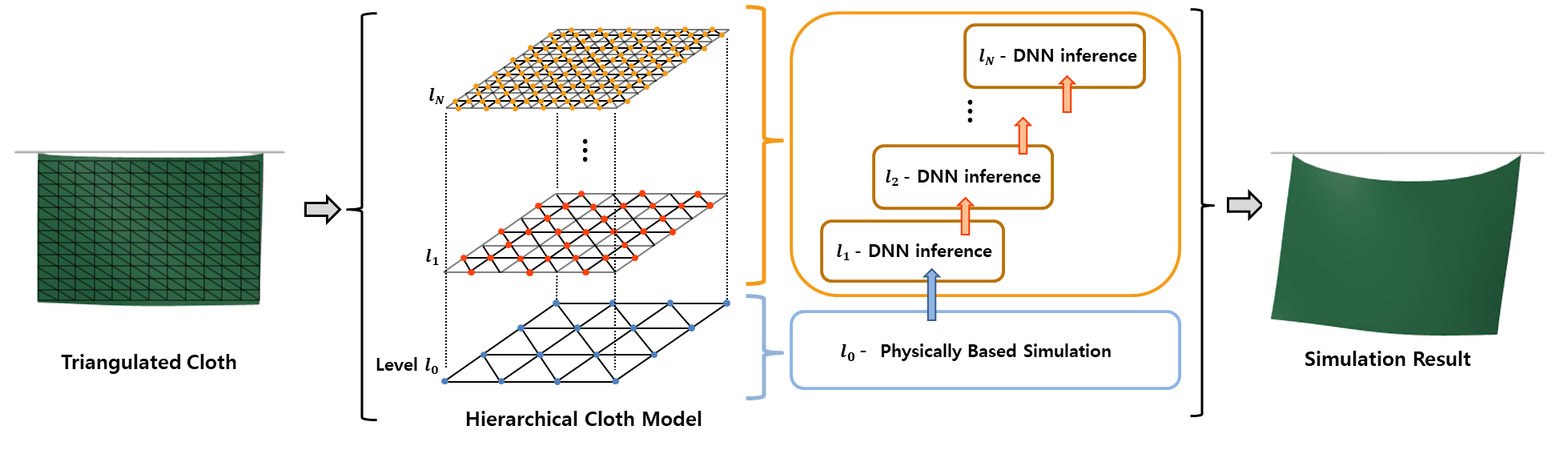}
\caption{System overview. We generate fast and reliable cloth simulation results using a hybrid method, which uses the conventional physically-based simulation and deep neural networks concurrently.}
\label{fig:overview}
\end{figure*}

\subsection{Physically-based Simulation with Deep Neural Networks}
Recently, DNN has been employed in the study of physically-based simulation, because it is particularly good at approximating complex computation processes. For fast fluid simulation, the replacement of pressure projection processes with DNN has been proposed~\cite{CNNFluid2016,Yang:2016:DPM:3074919.3074946}. Furthermore, DNN has been used to synthesize high-resolution flow simulations~\cite{Chu:2017:DSS:3072959.3073643} or to add liquid splashes~\cite{Um:2017:mlflip} to improve the quality of low-resolution flow simulations. Unlike previous approaches, there have been recent studies in which DNN generates physical motions of rigid body objects and simple elastic models, without using any physically-based simulation method. A neural physics engine~\cite{chang2016compositional} has been proposed that can learn intuitive physics using a physical scene and the past states of objects, and then generate a physical motion sequence for simple objects. Similar to~\cite{chang2016compositional}, visual interaction networks~\cite{NIPS2017_7040} have been proposed that can predict the next positions of objects following a learning process using a physically-based simulation video. 
Previous studies have shown that employing DNN in physically-based simulations yields reasonable simulation results. However, most studies have focused on fluid and smoke simulations. In addition, some studies have not generated reliable simulation results, due to significant errors such as inaccurate collisions and awkward motions. Our method is the first to use DNN for cloth simulation, and employs both the conventional physically-based simulation method and DNN simultaneously for a reliable simulation.

\section{Hierarchical Cloth Simulation System Using Deep Neural Networks}
Hierarchical cloth simulation using DNN is a simulation method that incorporates physically-based simulation and the inference of DNN. Fig. \ref{fig:overview} illustrates the flow of our entire method. Before the simulation, our system constructs a hierarchy of the cloth model in the pre-processing stage. The coarsest level (${l}_{0}$) in the hierarchy is simulated using the conventional physically-based simulation method. Finer levels (${l}_{i}, i=1,...,N$) are generated using the inference of DNN models, based on the input data from the previous levels ${l}_{i-1}$. 

\begin{figure}[t]
\begin{center}
\includegraphics[width=2.3in]{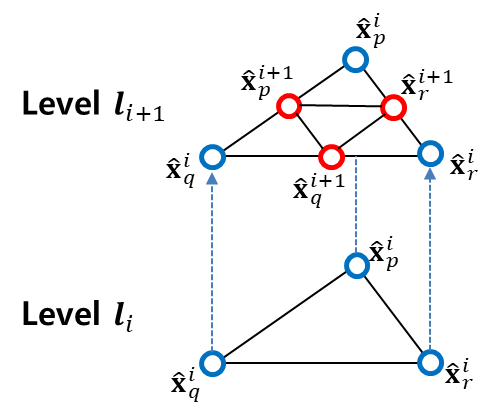}
\end{center}
\caption{Hierarchy of our cloth model. A triangle formed by three masses (blue dots) in level $l_i$ is subdivided into four triangles by adding three masses (red dots) in $l_{i+1}$ in the initial hierarchical cloth system.}
\label{fig:hierarchy}
\end{figure}

\subsection{Hierarchical Cloth Model}
In the initial hierarchical cloth system, triangles are recursively partitioned as the level of the hierarchy increases. A triangle formed by three masses ${\widehat{\mathbf{x}}}_{p}^{i}$, ${\widehat{\mathbf{x}}}_{q}^{i}$, and ${\widehat{\mathbf{x}}}_{r}^{i}$ at the level $i$ is subdivided into four triangles, by adding three masses ${\widehat{\mathbf{x}}}_{p}^{i+1}$, ${\widehat{\mathbf{x}}}_{q}^{i+1}$, and ${\widehat{\mathbf{x}}}_{r}^{i+1}$ at the level $i+1$. The added masses are the center points of each edge of the original triangle (see Fig. \ref{fig:hierarchy}). 

\subsection{Implicit Integration for Level ${l}_{0}$}
Consider the particles with positions $\mathbf{x}^{0}(t)$, velocities $\mathbf{v}^{0}(t)$, and the mass matrix $\mathbf{M}^{0}$ in the coarsest level ${l}_{0}$ of the hierarchical cloth model. The next positions $\mathbf{x}^{0}(t+\Delta t)$ of the particles in the simulation are calculated using implicit integration equations:
\begin{equation}
\mathbf{x}^{0}(t+\Delta t)=\mathbf{x}^{0}(t)+\mathbf{v}^{0}(t+\Delta t)\Delta t,
\end{equation}
\begin{equation}
\mathbf{M}^{0}\mathbf{v}^{0}(t+\Delta t)=\mathbf{M}^{0}\mathbf{v}^{0}(t) + \left ( \mathbf{f}_{\mathrm{ext}}(t)+ \mathbf{f}(t+\Delta t) \right )\Delta t,
\end{equation}
where $\mathbf{f}_{\mathrm{ext}}$ represents external forces, and $\mathbf{f}$ represents the implicit forces of the model. The calculation of Eq. (1) and (2) can be converted into the following minimization problem~\cite{martin2011example,gast2015optimization}:
\begin{equation}
\begin{split}
\mathbf{x}^{0}(t+\Delta t)=\textrm{arg}\min_{\mathbf{x}^{0}} & \left ( \frac{1}{2\Delta t^2} \left \| {\mathbf{M}^0}^\frac{1}{2} (\mathbf{x}^0(t) - \tilde{\mathbf{x}}^0(t+\Delta t)) \right \|^2 \right. \\
 & \left. + U(\mathbf{x}^0) \vphantom{\int_1^2} \right ),
\end{split}
\end{equation}
where $\tilde{\mathbf{x}}^0(t+\Delta t)$ is the predicted next position without $\mathbf{f}$, and $U$ is the sum of different energy terms that effect the cloth model. Eq. (3) can be solved using a method that applies the alternating direction method of multipliers (ADMM) algorithm~\cite{boyd2011distributed} for the implicit integration~\cite{Narain2016,overby2017admmpd}. In this paper, we use projective dynamics, applying ADMM~\cite{Narain2016} to calculate the next position of the coarsest level. 

\begin{figure}
\begin{center}
\includegraphics[width=3.4in]{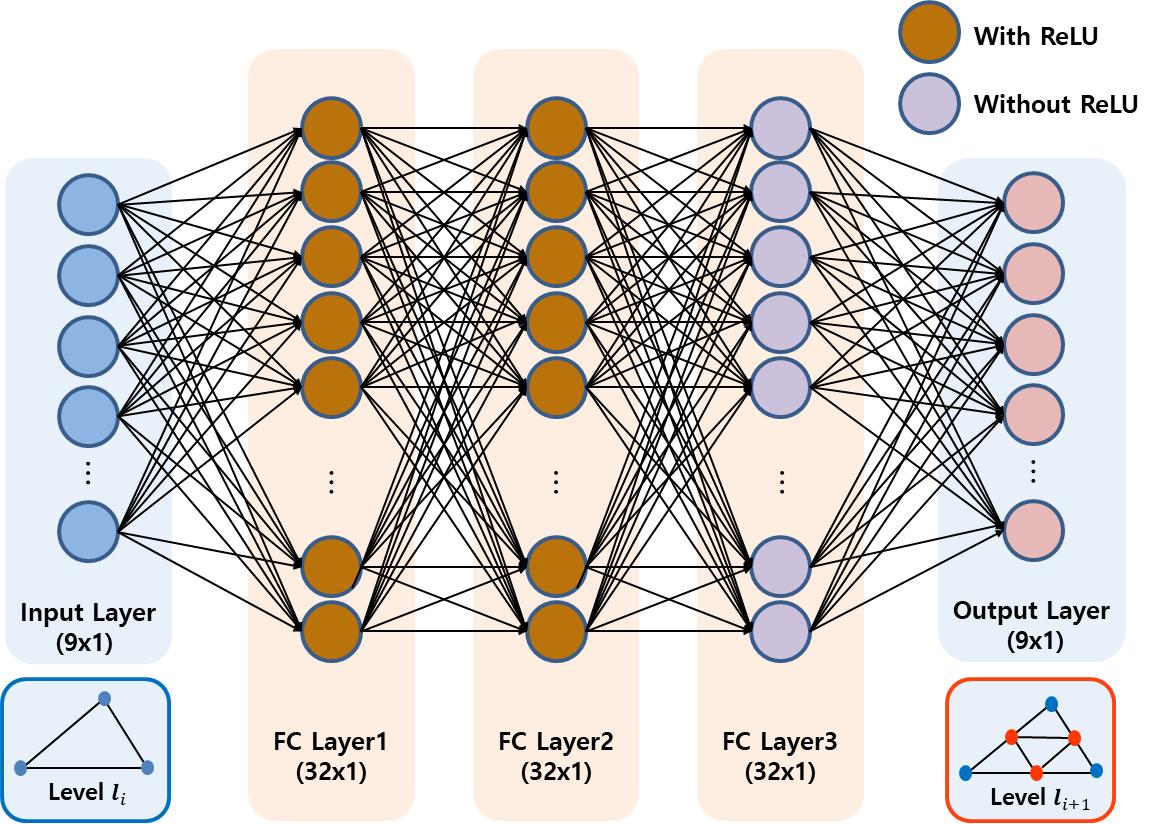}
\end{center}
\caption{DNN model for inference. The DNN model consists of three fully connected (FC) hidden layers and rectified linear unit activation (ReLU) functions.}
\label{fig:dnn}
\end{figure}

\subsection{Deep Neural Networks for Inference}
We generate the next positions of masses in finer levels by inferences of DNN models. 
Because an inference of a DNN model occurs in each triangle separately, the DNN model needs to be composed of simple neural networks. We use a fully connected (FC) layer and rectified linear unit (ReLU) activation function for constructing the DNN model. As described in Fig. \ref{fig:dnn}, the DNN model consists of three FC hidden layers and ReLU functions. The first and second FC layers calculate an intermediate vector of size 32, and transmit it to the next layer through a ReLU function. The third FC layer calculates the final out vector of size nine without a ReLU function. 

\begin{figure}[t]
\begin{center}
\includegraphics[width=3.5in]{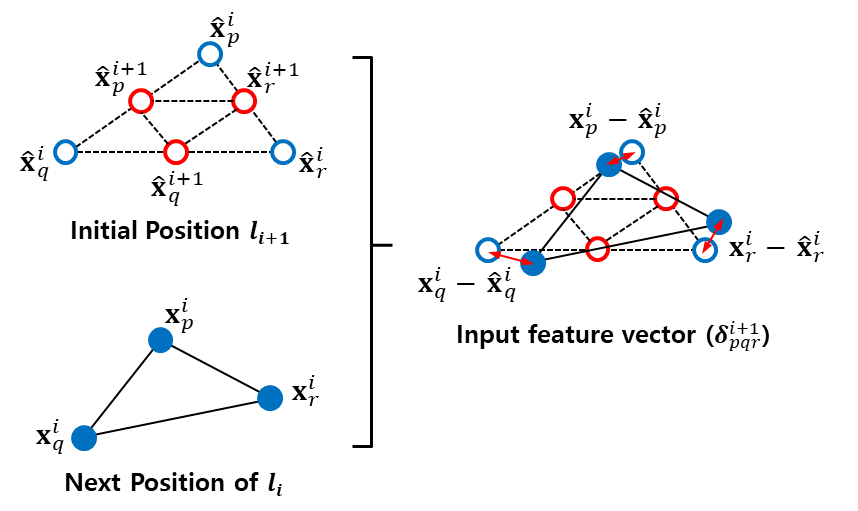}
\end{center}
\caption{Input feature vector for inference. The input feature vector of three masses in $l_{i+1}$ is calculated using the local differences compared to the initial position of the triangle in $l_i$.}
\label{fig:inputvec}
\end{figure}

\begin{figure}[t]
\begin{center}
\includegraphics[width=2.7in]{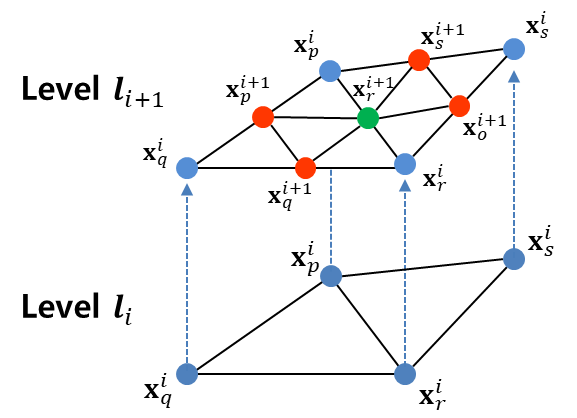}
\end{center}
\caption{Process for two different output vectors. A mass (green dot) has two output vectors from different inferences. The final output vector is calculated as the average of the two output vectors.}
\label{fig:overlap}
\end{figure}

\begin{algorithm}[t]
	\SetAlgoNoLine
	\footnotesize 
	{\bf Cloth quantities:} 
	\\ hierarchy of cloth model $l$, initial position $\widehat{\mathbf{x}}$,
	\\ current position $\mathbf{x}$, temporal position $\tilde{\mathbf{x}}$,
	\\ velocity $\mathbf{v}$, time step $\Delta t$, external force $\mathbf{f}_{\mathrm{ext}}$
	\\
	\vspace{5pt}
	{\bf Pre-computation:}\\
	\For {\bf all masses $i$}{
		Create hierarchy of cloth model and assign to level ($l_{0}, ... , l_{N}$)
	}
	\For {\bf all masses $i$ in $l_0$}{
		Create mass-spring and bending constraints 
	}
	\vspace{5pt}

	{\bf Runtime simulation:}\\ 
	// Accurate physically-based simulation\\
	\For {\bf all masses $i$ in $l_0$}{
		$\mathbf{v}_i \leftarrow \mathbf{v}_i+\mathbf{f}_{\mathbf{ext}}(\mathbf{x}_i) \Delta t $\\
		$\tilde{\mathbf{x}}_i \leftarrow \mathbf{x}_i+\mathbf{v}_i \Delta t $\\
	}
	\For {\bf all masses $i$ in $l_0$}{
		$\mathbf{x}_i$ $\leftarrow$ Projective Dynamics and Applying ADMM($\tilde{\mathbf{x}}_i$)\\
	}	
		
	\vspace{5pt}
	//Inference of DNN models\\
	\For {\bf $i=1$ to $N$}
	{
		\For {\bf all triangles $j$ in ${l}_{i}$}
		{
			$\delta_{j}^{i} \leftarrow$ Create Inputvector($\mathbf{x}_{j}^{i-1}$, $\widehat{\mathbf{x}}_{j}^{i-1}$) \\
			$\mathbf{o}_{j}^{i} \leftarrow$ Inference of DNN model($\delta_{j}^{i}$)\\
			$\mathbf{x}_{j}^{i} \leftarrow$ Update position($\mathbf{o}_{j}^{i}$)\\
		}
	}
	
	\caption{Hierarchical Cloth Simulation using Deep Neural Networks.}
	\label{alg:fullAlg}
\end{algorithm}

The input and output of the DNN model consist of the nine dimensional vectors that describe the local differences of the triangle compared to its initial position. An input feature vector is defined as
\begin{equation}
\mathbf{\delta}_{pqr}^{i+1}=\left [(\mathbf{x}_{p}^{i}-\widehat{\mathbf{x}}_{p}^{i}), (\mathbf{x}_{q}^{i}-\widehat{\mathbf{x}}_{q}^{i}), (\mathbf{x}_{r}^{i}-\widehat{\mathbf{x}}_{r}^{i})  \right ],
\end{equation}
for the inference of the three masses $\mathbf{x}_{p}^{i+1}$, $\mathbf{x}_{q}^{i+1}$, and $\mathbf{x}_{r}^{i+1}$ of the triangle in $l_{i+1}$, where $\widehat{\mathbf{x}}_{p}^{i}$, $\widehat{\mathbf{x}}_{q}^{i}$, and $\widehat{\mathbf{x}}_{r}^{i}$ are the initial positions of $\mathbf{x}_{p}^{i}$, $\mathbf{x}_{q}^{i}$, and $\mathbf{x}_{r}^{i}$ in $l_i$ (see Fig. \ref{fig:inputvec}). 
An output vector $\mathbf{o}_{pqr}^{i+1}$ is defined as
\begin{equation}
\mathbf{o}_{pqr}^{i+1}=\left [(\mathbf{x}_{p}^{i+1}-\widehat{\mathbf{x}}_{p}^{i+1}), (\mathbf{x}_{q}^{i+1}-\widehat{\mathbf{x}}_{q}^{i+1}), (\mathbf{x}_{r}^{i+1}-\widehat{\mathbf{x}}_{r}^{i+1})  \right ].
\end{equation}
The next positions of $\mathbf{x}_{p}^{i+1}$, $\mathbf{x}_{q}^{i+1}$, and $\mathbf{x}_{r}^{i+1}$ can be determined by adding their initial positions to $\mathbf{o}_{pqr}^{i+1}$: 
\begin{equation}
\left [\mathbf{x}_{p}^{i+1}, \mathbf{x}_{q}^{i+1}, \mathbf{x}_{r}^{i+1} \right ]=\left [\widehat{\mathbf{x}}_{p}^{i+1}, \widehat{\mathbf{x}}_{q}^{i+1}, \widehat{\mathbf{x}}_{r}^{i+1} \right ] + \mathbf{o}_{pqr}^{i+1}.
\end{equation}
Inferences of the DNN model belonging to the same level can proceed in parallel. When a mass has two output vectors from different DNN inferences (a green dot in Fig. ~\ref{fig:overlap}), the final output vector of the mass is calculated as the average of the two different output vectors. 

When the hierarchical cloth model has $N$ finer levels, we use $N$ DNN models for inference. Each DNN model is trained using the training data obtained at the target finer level, and is used only to generate the next position of the target finer level. In the training process for the DNN model, the loss function is defined as the root mean squared error (RMSE) between the ground-truth vector $\mathbf{g}$ and the output vector:
\begin{equation}
Loss = \sqrt{\frac{1}{n}\sum_{i=1}^{N}\sum_{j}(\mathbf{g}_{j}^{i}-\mathbf{o}_{j}^{i})^2},
\end{equation}
where $j$ represents the indices of triangles in $l_i$, and $n$ is the total number of inferences of the DNN models in finer levels. 

\subsection{Runtime Simulation Process}
Algorithm~\ref{alg:fullAlg} shows the pseudo-code for the hierarchical cloth simulation using DNN. In the pre-computation stage, our method builds a hierarchical cloth model and constructs mass-spring and bending constraints of $l_0$ for using the conventional physically-based simulation. In the runtime simulation stage, we first calculate the next positions of all masses in $l_0$, using the projective dynamics and applying ADMM. After updating the next positions of $l_0$, our method generates the next positions of finer levels using the inference of DNN models. 

\begin{table}[t]
  \caption{The loss values measured during the training and test process of the proposed DNN.}
  \label{tab:t1}
  \renewcommand{\arraystretch}{1.3}
\begin{tabular}{ccccc}
\hline
\multirow{2}{*}{DNN Model} & \multicolumn{4}{c}{Training epoch (times)}                              \\ \cline{2-5} 
                           & \multicolumn{1}{c}{100} & \multicolumn{1}{c}{1500} & 3000     & 5000     \\ \hline
3*FC-32 for $l_1$          & 1.48e-05               & 8.14e-06                  & 2.28e-06 & 1.63e-06 \\  
3*FC-32 for $l_2$          & 2.94e-04               & 1.45e-06                  & 3.13e-07 & 2.62e-07  \\ \cline{1-5} 
\end{tabular}
\end{table}

\begin{figure}[t]
\begin{center}
\includegraphics[width=3.3in]{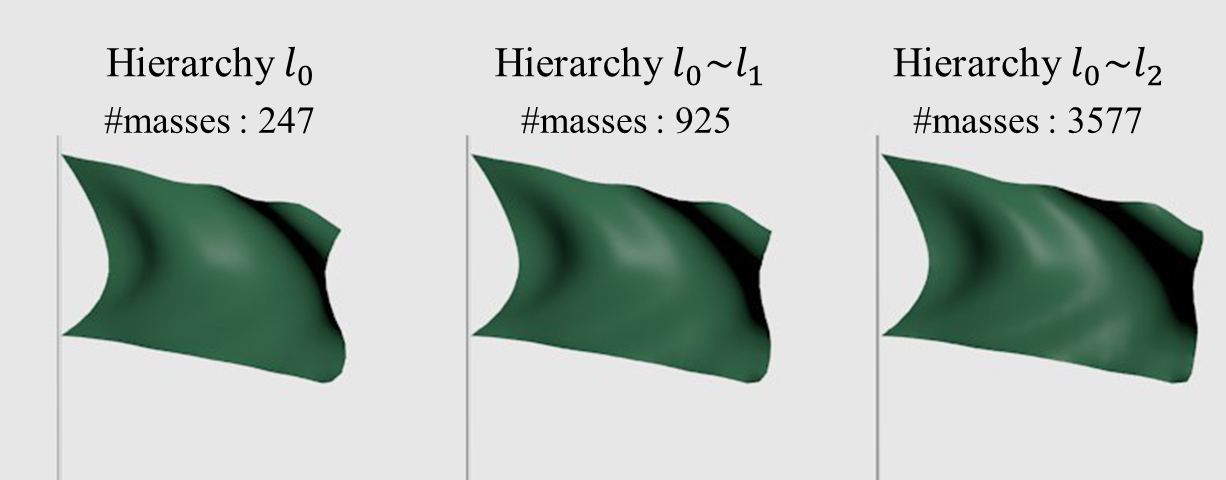}
\end{center}
\caption{Simulation results on different hierarchy levels. The result of calculating $l_0$ using the conventional method (left) and using DNN models to infer $l_1$ (middle) and $l_2$ (right).}
\label{fig:res_diff_levels}
\end{figure}

\section{Results}
We implemented the hierarchical cloth simulation system by integrating the physically-based simulation using projective dynamics and applying ADMM~\cite{Narain2016} and our DNN models. Because we used a cloth model that has three hierarchies (${l}_{i}, i=0,1,2$) to generate simulation results, we constructed two DNN models for $l_1$ and $l_2$.  
For the training data on each DNN model, we collected various cloth simulation results using the projective dynamics and applying the ADMM method while changing the cloth resolution and cloth material parameter values, such as the stiffness of the mass-spring and bending constraints. We sampled 2,873,465 and 3,012,009 input feature vector and ground truth pairs for the training data for the DNN models $l_1$ and $l_2$, respectively, from the collected simulation results. We constructed the DNN models using TensorFlow~\cite{tensorflow2015-whitepaper}, and trained them in an environment equipped with Intel Core i7-3770, 32 GB main memory, and NVIDA GTX 970 graphic card. The weights and biases of the DNN model were updated through back propagation, using the Adam optimizer~\cite{DBLP:journals/corr/KingmaB14}.
Table~\ref{tab:t1} shows the loss values of DNN models measured during the training process. As the table shows, the loss sharply decreased as the training was repeated. All the results presented in this paper were generated using DNN models trained through 5,000 training epochs. When applying our method to generate simulation results, the physically-based simulation and inference of DNN employed parallel programming using OpenMP~\cite{Dagum:1998:OIA:615255.615542}. 

\begin{figure}[t]
\begin{center}
\includegraphics[width=3.3in]{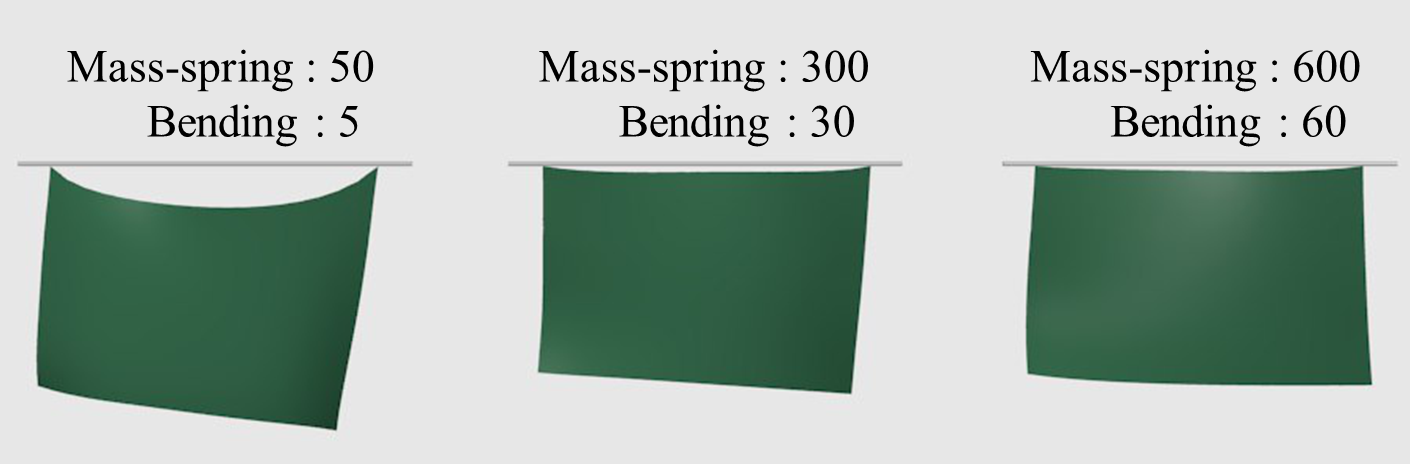}
\end{center}
\caption{The results of changing the stiffness values of the mass-spring and bending constraints. }
\label{fig:res_diff_stiffness}
\end{figure}

\begin{figure}[t]
\begin{center}
\includegraphics[width=3.3in]{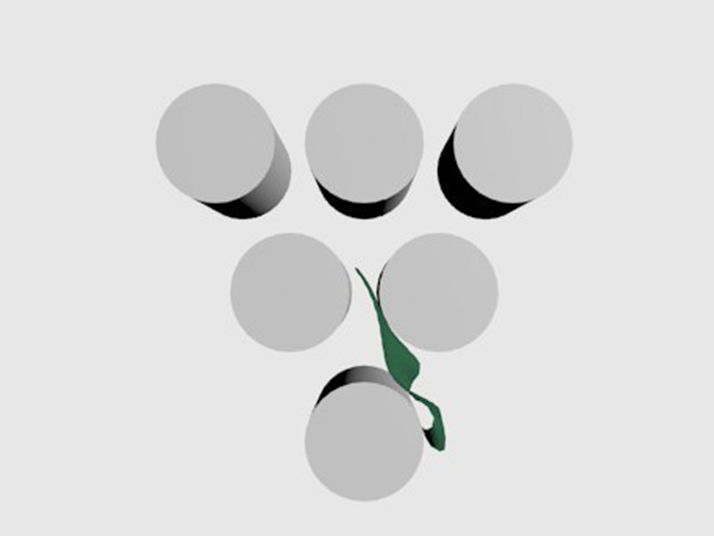}
\end{center}
\caption{The result when there are collisions between the cloth and external objects.}
\label{fig:res_collision}
\end{figure}

\subsection{Simulation Results}
The three flags shown in Fig.~\ref{fig:res_diff_levels} illustrate the simulation results for our method. The flag on the left is the result of $l_0$ calculated using the conventional physically-based simulation method. The middle and right flags are the results inferred up to $l_1$ and $l_2$, respectively, using DNN models. As the results show, our method can generate reliable and high-quality simulation results as the hierarchy rises. 
Because we used the training data of various conditions, it is inferred that our method generates stable simulation results even when the simulation conditions are varied. Fig.~\ref{fig:res_diff_stiffness} presents the results when changing the stiffness values of the mass-spring and bending constraints of the cloth model. Fig.~\ref{fig:res_collision} shows the result when there are collisions between the cloth and external objects. Even under various conditions, our method can generate reliable simulation results. 

\subsection{Comparisons with Conventional Methods}
\begin{figure}[t]
\begin{center}
\includegraphics[width=3.4in]{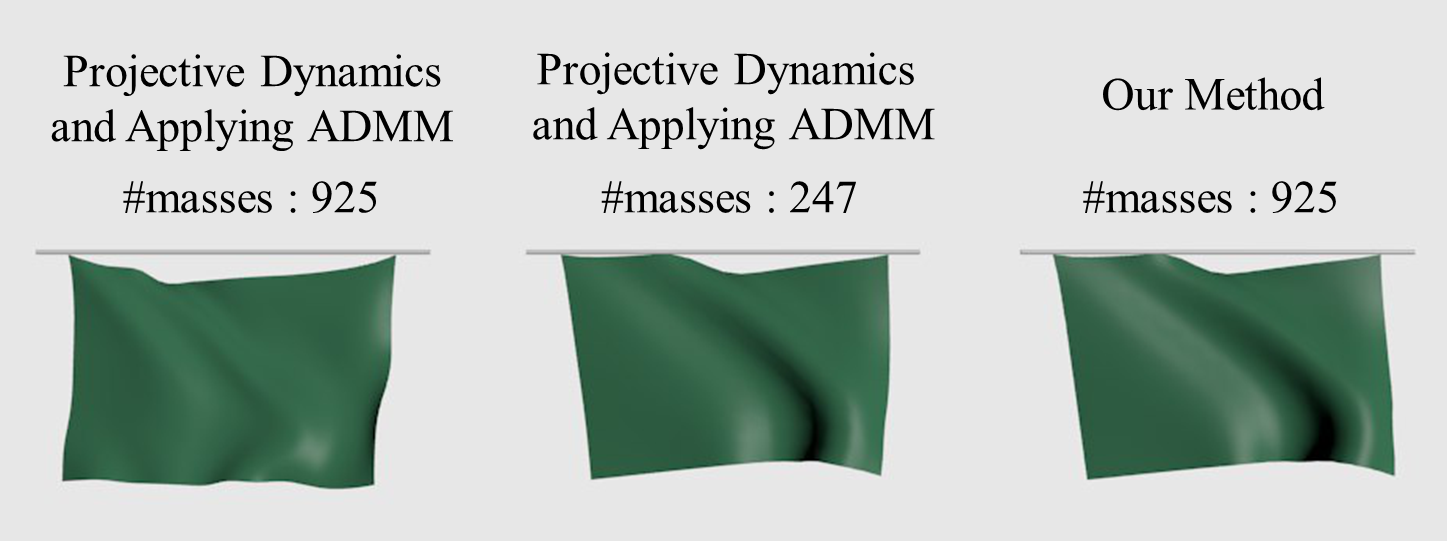}
\end{center}
\caption{The cloth simulation results for high-resolution (left) and low-resolution (middle) calculated using the conventional method, and high-resolution (right) generated using our method.}
\label{fig:compwithothers}
\end{figure}

We compared the simulation results and computational performance with the conventional physically-based simulation method. 
The left and middle images in Fig.~\ref{fig:compwithothers} show the high-resolution and the low-resolution results calculated using projective dynamics and applying the ADMM method~\cite{Narain2016}. The right image in Fig.~\ref{fig:compwithothers} is the high-resolution result generated using our method.
As the left and the right images show, our simulation result is different to those of the conventional physically-based simulation method within the same resolution. Because our method calculates only the coarsest level using the conventional physically-based simulation method, our result is not as accurate as that of the conventional physically-based simulation. 
However, as the middle and the right images show, our simulation result is similar to the low-resolution simulation calculated using the conventional physically-based simulation method. This is because our DNN models infer the next positions of finer levels based on the result of the coarsest level. 

\begin{table}[t]
\centering
\caption{Comparison of time performance between conventional methods and our method.}
\label{tab:t2}
\renewcommand{\arraystretch}{1.2}
\begin{tabular}{ccccc}
\hline
Exmaple & \#masses & \begin{tabular}[c]{@{}c@{}}CG\\(ms)\end{tabular} & \begin{tabular}[c]{@{}c@{}}ADMM\\(ms)\end{tabular} & \begin{tabular}[c]{@{}c@{}}Our method\\ (ms)\end{tabular} \\ \hline
Flag    & 3577   & 87.5                                                  & 76.4                                                & 33.8                                                      \\ 
Cloth   & 925   & 21.3                                                  & 19.2 & 15.4   \\ 
\hline
\end{tabular}
\end{table}

Table~\ref{tab:t2} compares the time performance for examples with the same resolution calculated using implicit integration~\cite{baraff1998large} with the conjugate gradient (CG) method, projective dynamics and applying ADMM~\cite{Narain2016}, and our method. 
In the experiment, by fixing the number of iterations of the CG and ADMM methods, we maintained the simulation results calculated by the two conventional methods with similar errors. The number of iterations for the CG and ADMM methods were 100 and 20, respectively. For our method, we applied the same number of iterations of the ADMM method for the coarsest level, and generated finer levels using the inference of DNN models. As the table shows, our method generates simulation results faster than the conventional physically-based simulation methods.
As described in Fig.~\ref{fig:compwithothers}, our method is not as accurate as the conventional physically-based simulation method, but it is fast and appropriate for generating high-quality simulation results that are as reliable as the conventional physically-based simulation.

\begin{figure}[t]
\begin{center}
\includegraphics[width=3.3in]{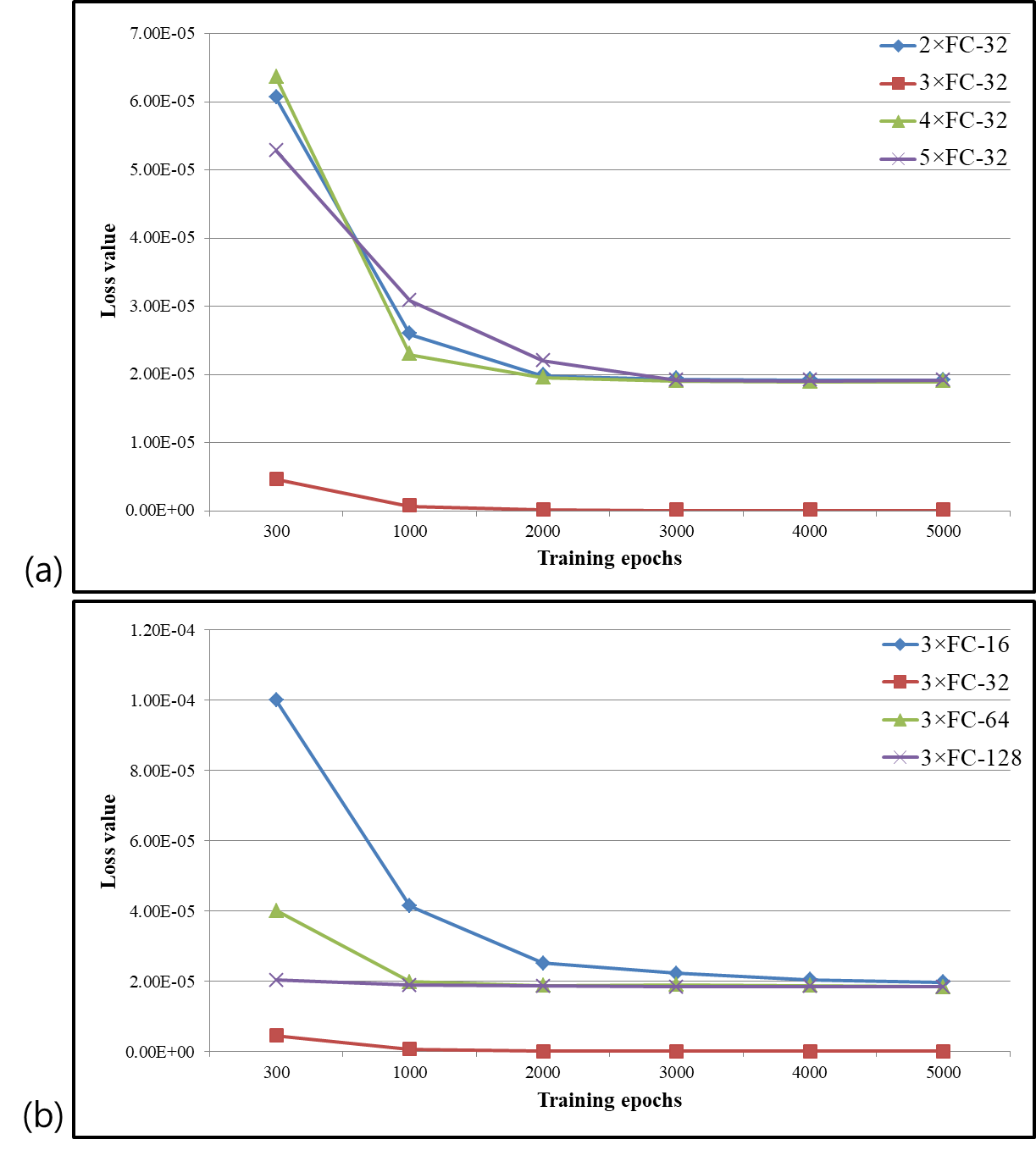}
\end{center}
\caption{Performance comparison of DNN models composed of various structures: (a) DNN models composed of two, three, four, and five FC hidden layers. (b) DNN models using FC layers of size 16, 32, 64, and 128, respectively.}
\label{fig:different_dnn_structures}
\end{figure}

\subsection{Performance Comparison According to DNN Structure}
We compared the performance of the DNN model composed of various structures, to determine the most suitable DNN model for the inference. Fig.~\ref{fig:different_dnn_structures} (a) shows the result of measuring the loss values as the training epoch increases when the DNN model consists of two, three, four, and five FC hidden layers, respectively. The size of the single FC layer used in the experiment was fixed at 32. As the graph shows, the loss value is the smallest in the model with three FC layers, in which case the loss value decreases at the fastest rate as the number of training epochs increases. Fig.~\ref{fig:different_dnn_structures} (b) shows the result of measuring the loss values as the number of training epochs increases, when the DNN model uses FC layers of size 16, 32, 64, and 128, respectively. The DNN model used in this experiment consisted of three FC layers. As the graph shows, the loss value is the smallest in the model using FC layers of size 32. 

\section{Conclusion}
We have proposed a hybrid cloth simulation method using both the conventional physically-based simulation method and deep neural networks (DNN). The proposed method calculates the next position of the coarsest level using an accurate physically-based simulation method, to guarantee reliable results. For finer levels in the hierarchical model, a simple DNN with a low-cost inference time is adopted in the inference process, for the fast generation of simulation results. We compared the performance and simulation results of our method with those of conventional physically-based simulation methods.
The results generated by our method are not as accurate as those of conventional physically-based simulations, but our method is fast and appropriate for generating high quality results that are as reliable as those of conventional physically-based simulations. As a result, we have shown that our method is suitable for computer graphics contents that require a fast and reliable physically-based simulation method.

\begin{figure}[t]
\begin{center}
\includegraphics[width=3.3in]{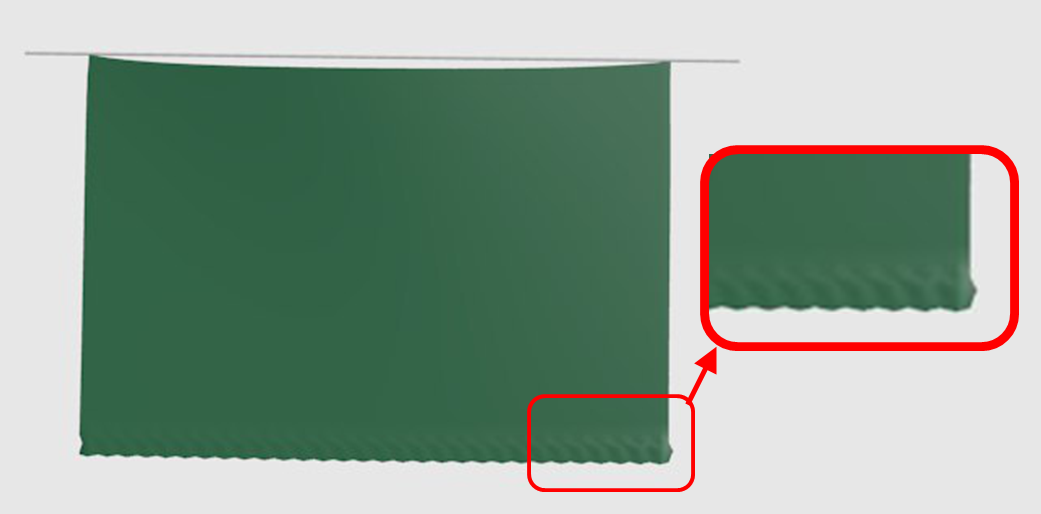}
\end{center}
\caption{Failure case. When the cloth is over-stretched, unusual wrinkles form in the lower part of the cloth due to inaccurate inferences of DNN.}
\label{fig:failure}
\end{figure}

Fig.~\ref{fig:failure} shows a simulation result when the cloth is over-stretched by external forces. Because of inaccurate inferences of the DNN models, unusual wrinkles form in the lower part of the cloth, and these do not disappear even if the simulation is repeated. As the figure shows, because DNN models are trained from a pre-computed result using the conventional physically-based simulation method, our method is unable to generate reliable results for unexpected situations that are not included in the pre-computed results. 
As shown in Fig.~\ref{fig:compwithothers}, our simulation generates results that differ from those calculated using the conventional physically-based simulation method. Therefore, our method is limited to use in fields requiring accurate simulation results.
We plan to extend the proposed method, and apply it to the simulation of fluids and other kinds of deformable objects. As our results show, in other areas it can be expected that fast and reliable simulations will be possible by using DNN with a conventional physically-based simulation method. In addition, different types of DNN~\cite{NIPS2014_5423,CycleGAN2017} with an improved performance could be used for inference. We plan to employ them for enhancing the quality of simulation results. 


\bibliographystyle{spmpsci}
\bibliography{h_dnn} 

\end{document}